\documentclass[aps,prl,twocolumn,showpacs,floatfix,twoside,superscriptaddress]{revtex4-2}
\usepackage{amssymb}
\usepackage{graphicx}
\usepackage{amsmath}
\usepackage{amsfonts}
\usepackage{bbold}
\usepackage{esint}
\usepackage{color}
\usepackage[dvipsnames]{xcolor}
\usepackage{float}
\usepackage{verbatim}
\usepackage[colorlinks=true,citecolor=blue,linkcolor=blue,urlcolor=blue]{hyperref}
\usepackage[justification=centerlast,font={small}]{caption}

\DeclareMathOperator{\im}{Im}
\DeclareMathOperator{\re}{Re}

\DeclareMathOperator{\sign}{sign}

\newcommand{\BT}{\mathcal{T}}

\newcommand{\BP}{\mathcal{P}}

\newcommand{\BU}{\mathcal{U}}

\begin{document}
 
 \title{Interplay of pseudo-Hermitian symmetries and degenerate manifolds in the eigenspectrum of non-Hermitian systems
 }
\author{Grigory A. Starkov}
\affiliation{Institute for Theoretical Physics and Astrophysics,
University of Würzburg, D-97074 Würzburg, Germany}
\email{grigorii.starkov@uni-wuerzburg.de}

\date{\today}

\begin{abstract}
In this letter, we study how the spectrum of pseudo-Hermitian systems is influenced by the ambiguity in the choice of the pseudo-metric operator.
In particular, we analyze the case when different parameter-independent choices of pseudo-metric are possible and how it can lead to the appearance of robust degenerate manifolds in the parameter space of the system.
\end{abstract}

\maketitle

\begin{figure*}[t]
\begin{center}
\includegraphics[width=500pt]{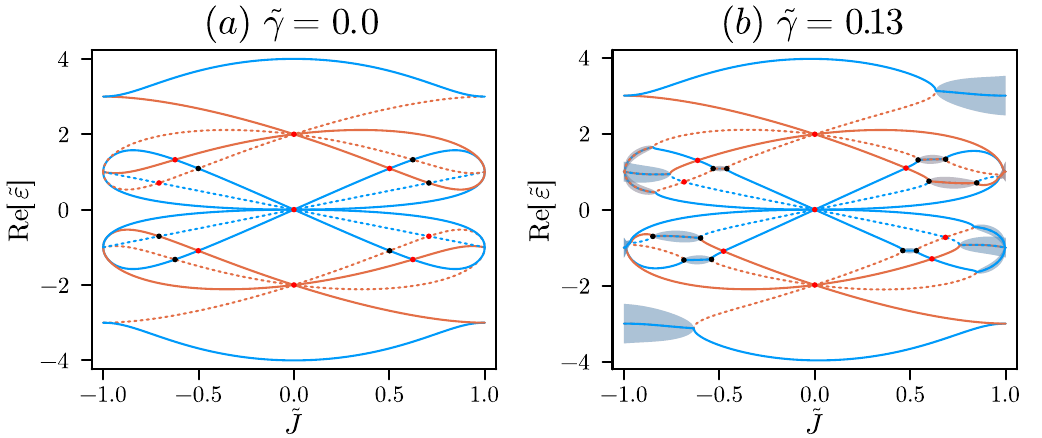}
\caption{Eigenspectrum of the non-Hermitian transverse-field Ising chain in the case of longitudinal gain (loss). The panels plot the real parts of the normalized eigenenergies $\re[\tilde\varepsilon]$ while the imaginary parts are displayed as shaded ribbons with the width proportional to $\im[\tilde\varepsilon]$. The solid (dashed) linestyle denotes the $\zeta_\BP=+1$ ($\zeta_\BP=-1$) topological index with respect to the first choice of the pseudo-metric operator. The blue (orange) color denotes the $\eta=+1$ ($\eta=-1$) topological index with respect to the second choice of the pseudo-metric operator.  Diabolical points protected by the combination of two pseudo-metric operators (red dots) are stable.
In comparison, the crossings of the levels that are allowed to form second-order \textit{EP}s by all the choices of pseudo-metric operator (black dots in panel $(a)$) naturally split into pairs of second-order \textit{EP}s (pairs of black dots in panel $(b)$) as soon as gain (loss) is turned on.}
\label{figz}
\end{center}
\end{figure*}

\begin{figure*}[t]
\begin{center}
\includegraphics[width=500pt]{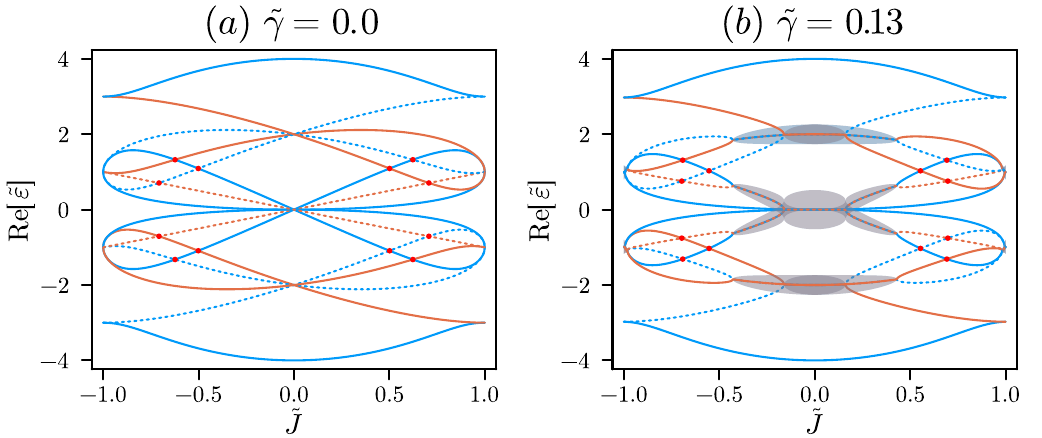}
\caption{Eigenspectrum of the non-Hermitian transverse-field Ising model in the case of transversal gain (loss). The panels plot the real parts of the normalized eigenenergies $\re[\tilde\varepsilon]$ while the imaginary parts are displayed as shaded ribbons with the width proportional to $\im[\tilde\varepsilon]$. The solid (dashed) linestyle denotes the $\zeta_\BP=+1$ ($\zeta_\BP=-1$) topological index with respect to the first choice of the pseudo-metric operator. The blue (orange) color denotes the $\eta=+1$ ($\eta=-1$) topological index with respect to the second choice of the pseudo-metric operator. Diabolical points protected by the combination of two pseudo-metric operators (red dots) are stable.}
\label{figx}
\end{center}
\end{figure*}

\begin{figure*}[t]
\begin{center}
\includegraphics[width=500pt]{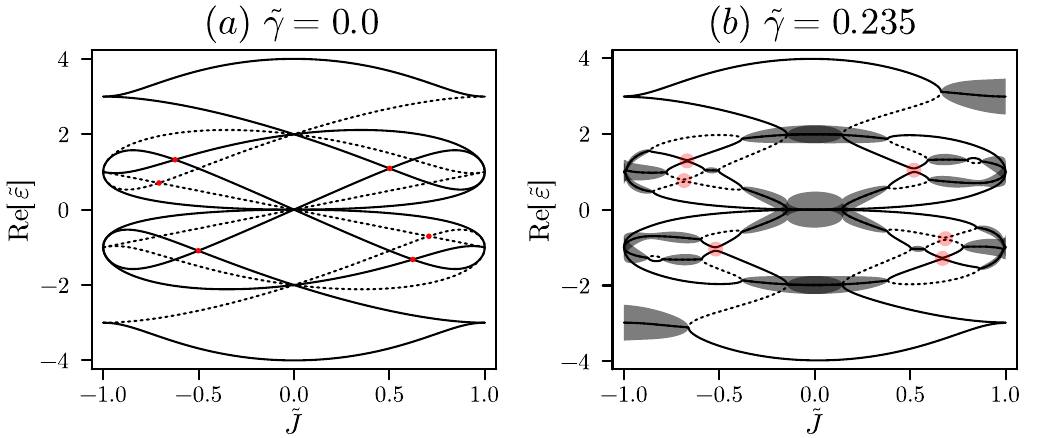}
\caption{Eigenspectrum of the non-Hermitian transverse-field Ising model in the case of mixed longitudinal-transversal gain (loss).
The panels plot the real parts of the normalized eigenenergies $\re[\tilde\varepsilon]$ while the imaginary parts are displayed as shaded ribbons with the width proportional to $\im[\tilde\varepsilon]$. There is only one parameter-independent choice of the pseudo-metric operator and the corresponding topological indices are denoted by solid ($\zeta_\BP=+1$) or dashed ($\zeta_\BP=-1$) linestyle.
Turning on gain (loss) transforms the same parity diabolical points (red dots in panel $(a)$) into avoided crossings (red shaded areas in panel $(b)$).}
\label{figxz}
\end{center}
\end{figure*}

\textit{Introduction\/.}--- Recent years show an increased interest in the study of non-Hermitian phenomena.
Non-Hermitian Hamiltonians arise ubiquitously as the effective description of open systems~\cite{Ashida-20} and generally in many-body interacting systems~\cite{Zyuzin-18, Fu-19, Budich-21, Kozii-17, Yoshida-18, Kazuhiro-19, Yoshida-20, Nagai-20, Michishita-20, Rausch-21, Lehmann-21, Crippa-21, Reitner-23}, where the system itself acts as a thermalizing bath for its constituents.
One of their most captivating features is the occurence of the special type of degeneracies called Exceptional Points (\textit{EP}s), where the coalescence of several eigenvalues is accompanied also by the simultaneous coalescence of the corresponding eigenvectors. The applications of \textit{EP}s include increasing sensitivity of quantum sensors~\cite{Wiersig-14, Wiersig-16, WiersigYang-17,  Khajavikhan-17, Khajavikhan-19}, 
mode switching in waveguides~\cite{Chong-16, Rotter-16, Chan-18, Ghosh-18} and optical microcavities~\cite{Ghosh-17, Ghosh-17b, Ghosh-19}, state conversion~\cite{Ghosh-20, Ghosh-21} and  laser emission management~\cite{Rotter-14, XZhang-16}, to name a few.
Despite promising practical properties, the realization of~\textit{EP}s is complicated by the necessity of tuning a system to the spectral vicinity of a required \textit{EP}. Nevertheless, the number of fine-tuning parameters can be lowered in the presence of symmetries. As such, investigation of the influence of the symmetries on the eigenspectrum of a non-Hermitian system is of immense practical interest.
In this regards, a widely considered class of symmetries is that of anti-Unitary symmetries~\cite{Hatsugai-21, Bergholz-21b, Sayyad-22, Kunst-22}, which all can be related either to $\BP\BT$-symmetry or to pseudo-Hermiticity~\cite{Hatsugai-21}. In addition to that, $\BP\BT$-symmetric systems are known to be pseudo-Hermitian~\cite{mostafazadeh2002pseudo,mostafazadeh2,mostafazadeh3,mostafazadeh2010, Ashida-20,ZhangQinXiao-20}, therefore we focus exclusively on the latter property.

A system is said to be pseudo-Hermitian when its Hamiltonian satisfies
\begin{equation}
    \hat\zeta \hat H \hat \zeta^{-1} = \hat H^\dagger,
\end{equation}
where $\hat\zeta$ is an invertible Hermitian operator called pseudo-metric.
The eigenvalues of a pseudo-Hermitian Hamiltonian can be either real or form complex conjugated pairs, and \textit{EP}s separate the corresponding parameter regions. In general, the pseudo-metric is not uniquely defined~\cite{bian2020conserved,agarwal2022conserved} and may depend on the parameters of the system~\cite{mostafazadeh2002pseudo,mostafazadeh2,mostafazadeh3,mostafazadeh2010, Gong-13, Deffner-15,Zhang-19a, Zhang-19b}, however, in a lot of relevant situations, one can choose it to be parameter-independent~\cite{bender1998real, ruter2010observation, Song-14, tetling2022linear, StarkovFistulEremin-22}.
In the latter case, in the parameter regions with real eigenvalues, the levels can be uniquely characterized by the topological indices $\pm1$
, which govern the formation of \textit{EP}s~\cite{StarkovFistulEremin-23, Krein-1950,GelfandLidskii-1955, StarzhinskiiYakubovich-1975, Melkani-23}: Second-order \textit{EP}s, where two levels coalesce, are provided only by the level pairs with opposite indices. The structure of higher-order \textit{EP}s is restricted insofar they can be obtained from several second-order \textit{EP}s tuned to coalesce.

Notably the parameter-independent choice of the pseudo-metric can itself be non-unique, which enables one to define the set of topological indices with respect to either of the pseudo-metrics. Taking into account the interplay between different sets of indices is important to predict the formation of \textit{EP}s. For example,  a second-order \textit{EP}s is realized only when it is allowed by all sets of topological indices. Moreover, such interplay can influence even normal non-defective degeneracies (also called diabolical points) which are present in non-Hermitian systems as well. The symmetry stabilization of the diabolical points has been studied recently in Refs.~\cite{Xue-20,Sayyad-22st}. Here, we discuss a completely new mechanism for such stabilization: formation of a second-order \textit{EP} can be allowed by one set of indices and forbidden by another one; in this case, a non-defective crossing of two levels is promoted to a robust degenerate manifold in the parameter space.


\textit{Pseudo-metric and the structure of eigenvectors\/.} --- Away from \textit{EP}s, the Hamiltonian has a complete biorthonormal basis of right and left eigenvectors~\cite{mostafazadeh2010, Ashida-20}:
\begin{align}
 \hat H(\vec p) |R_n(\vec p)\rangle & = \varepsilon_(\vec p) |R_n(\vec p)\rangle,\\
 \langle L_n(\vec p) | \hat H(\vec p) & = \varepsilon_n(\vec p)\langle L_n(\vec p)|,\label{left-right-eigvecs}
\end{align}
\begin{equation}
 \langle L_n(\vec p) | R_{n^\prime}(\vec p) \rangle = \delta_{n,n^\prime}.\label{completeness}
\end{equation}
We focus on the levels corresponding to the real eigenvalues.
In this case, the pseudo-metric is known to bijectively map the right eigensubspace onto the left one. For a level corresponding to a non-degenerate eigenvalue, it implies
\begin{equation}
    \hat \zeta |R_n(\vec p)\rangle = c_n(\vec p) |L_n(\vec p)\rangle.\label{mapping}
\end{equation}
For a degenerate eigenvalue, Eq.~\eqref{mapping} still holds provided we choose the right eigenvectors to diagonalize $\langle R_m|\hat \zeta | R_n\rangle$.

The biorthonormality condition~\eqref{completeness} does not completely fix the normalization of the left and right eigenvectors. We can use it to rescale both eigenvectors in such a way, as to remove $c_n(\vec p)$ almost completely with the exception of its sign:
\begin{equation}
\hat \zeta |R_n(\vec p)\rangle =\zeta_n |L_n(\vec p)\rangle,
\end{equation}
\begin{equation}
\zeta_n = \sign{\left[\sqrt{\langle R_n(\vec p)|\hat \zeta | R_n(\vec p)\rangle}\right]} = \pm1.\label{index-construction}
\end{equation}
The scalar product inside the sign turns zero only at the \textit{EP}s~\cite{Ashida-20,StarkovFistulEremin-23}. Therefore, the topological index $\zeta_n$ of the specific level is conserved in the whole parameter region where the corresponding eigenvalue stays real~\cite{StarkovFistulEremin-23}.

If there is more than one parameter-independent pseudo-metric operator, than one can define the topological indices with respect to either of them.
There is, however, an important caveat, that the right choice of the right and left eigenvectors satisfying Eq.~\eqref{mapping} could be different for different pseudo-metric operators. However, this problem can only arise for the left and right eigenvectors corresponding to a degenerate eigenvalue.\bigskip


\textit{Effective projected Hamiltonian in the vicinity of a degeneracy.} --- In the vicinity of a second-order degeneracy, be it defective or non-defective, we can project onto the states involved. We can use then the identity~\eqref{index-construction} to restrict the structure of the projected Hamiltonian.

Let the vector of parameters $\vec p^\prime$ correspond to a non-degenerate point in the vicinity of a second-order degeneracy, where the two levels involved have real eigenenergies. We can linearize the Hamiltonian around $\vec p^\prime$,
\begin{equation}
    \hat H(\vec p^\prime + \Delta \vec p) = \hat H (\vec p^\prime) +  \vec\nabla \hat H(\vec  p^\prime)\cdot\Delta\vec p ,
\end{equation}
and then project it onto the left $\langle L_i(\vec p^\prime)|$ and right $|R_i(\vec p^\prime)\rangle$ eigenvectors $i=1,2$ corresponding to the degeneracy. Assuming that the eigenvectors are rescaled to make Eq.~\eqref{index-construction} true, one can easily prove~\cite{StarkovFistulEremin-23} that the matrix elements of the projected Hamiltonian satisfy
\begin{equation}
    \langle L_i(\vec p^\prime)|\hat H(\vec p)|R_j(\vec p^\prime)\rangle = \zeta_i\zeta_j \overline{\langle L_j(\vec p^\prime)|\hat H(\vec p)|R_i(\vec p^\prime)\rangle},
\end{equation}
where the bar on top denotes the complex conjugation.
This identity restricts the structure of the projected Hamiltonian to
\begin{multline}
   H^P_{i,j}=\langle L_i(\vec p^\prime)|\hat H(\vec p)|R_j(\vec p^\prime)\rangle = \\
   \begin{pmatrix}
   \varepsilon_1(\vec p^\prime) + \vec u_1(\vec p^\prime)\cdot \Delta\vec p &\vec w(\vec p^\prime)\cdot \Delta \vec p\\
   \zeta_1\zeta_2  \vec w^*(\vec p^\prime) \cdot \Delta\vec p & \varepsilon_2(\vec p^\prime) + \vec u_2(\vec p^\prime)\cdot \Delta\vec p
   \end{pmatrix},\label{hproj}
\end{multline}
where $\vec u_1$ and $\vec u_2$ are real-valued.
As one can clearly see, the projected Hamiltonian can have a non-Hermitian form only if $\zeta_1\zeta_2=-1$, which means that the second-order \textit{EP}s are formed only by levels with opposite topological indices.

Let us assume now, that we have another choice of the pseudo-metric operator $\hat {\eta}$, which defines a compatible set of topological indices. We can repeat the same steps to derive the projected Hamiltonian in the basis of eigenvectors that satisfies Eq.~\eqref{index-construction} for this second operator. The new eigenvectors $\langle \tilde L_i(\vec p^\prime) | $ and $|\tilde R_i(\vec p^\prime)\rangle$ are related to the old ones by a simple rescaling that preserves the biorthonormality~\eqref{completeness}:
\begin{equation}
    |\tilde R_i(\vec p^\prime)\rangle = \alpha_i |R_i(\vec p^\prime)\rangle,\quad \langle \tilde L_i(\vec p^\prime) | = \frac{1}{\alpha_i} \langle L_i(\vec p^\prime) |.\label{rescaling}
\end{equation}
The projected Hamiltonian in the new basis is
\begin{multline}
    \tilde H^{P}_{i,j} = \langle \tilde L_i(\vec p^\prime)|\hat H(\vec p)|\tilde R_j(\vec p^\prime)\rangle = \\
    \begin{pmatrix}
   \varepsilon_1(\vec p^\prime) + \vec {\tilde u}_1(\vec p^\prime)\cdot \Delta\vec p &\vec {\tilde w}(\vec p^\prime)\cdot \Delta \vec p\\
   \eta_1\eta_2  \vec {\tilde w}^*(\vec p^\prime) \cdot \Delta\vec p & \varepsilon_2(\vec p^\prime) + \vec {\tilde u}_2(\vec p^\prime)\cdot \Delta\vec p
   \end{pmatrix}.\label{hproj2}
\end{multline}

The two forms of the projected Hamiltonian~\eqref{hproj2} and~\eqref{hproj} should lead to the same eigenspectrum, i.e., they are similar matrices with identical characteristic polynomials.
This means that the products of off-diagonal elements, equal up to a sign to the free term of the characteristic polynomial, should be the same for two matrices:
\begin{equation}
\zeta_1\zeta_2 |\vec w(\vec p^\prime)\cdot \Delta\vec p|^2 = \eta_1\eta_2 |\vec {\tilde w}(\vec p^\prime)\cdot \Delta\vec p|^2.
\end{equation}
If now
\begin{equation}
\zeta_1\zeta_2=-\eta_1\eta_2,\label{zero-cond}
\end{equation}
then the off-diagonal matrix element is forced to be zero in either of the bases: $\vec w(\vec p^\prime) = \vec {\tilde w}(\vec p^\prime) = 0$.\bigskip

\textit{Robust degenerate manifold\/.} ---
Let us now choose the expansion point $\vec p^\prime$ corresponding to a second-order diabolical point, i.e. second order non-defective degeneracy. If the indices of the two levels satisfy Eq.~\eqref{zero-cond},
we can rewrite the projected Hamiltonian~\eqref{hproj} as
\begin{equation}
H^P = \left[\varepsilon(\vec p^\prime) + \vec u_+ \cdot \Delta\vec p\right]\mathbb{1} + (\vec u_- \cdot \Delta\vec p)\sigma_z.
\end{equation}
Here, $\mathbb{1}$ and $\sigma_z$ are the identity and the $z$ Pauli $2\times2$ matrices.
The vectors $\vec u_\pm$ are symmetric and antisymmetric combinations of $\vec u_{1,2}$:
\begin{equation}
\vec u_{\pm} = \frac{\vec u_1 \pm \vec u_2}{2}.
\end{equation}
The equation $\vec u_-\cdot \Delta\vec p=0$ defines then locally a manifold of diabolical points. This manifold is robust in the sense that it must naturally appear as long as the dimensionality of the parameter space $\geqslant2$. Although the condition~\eqref{zero-cond} prevents the two levels either from opening the gap or from forming an \textit{EP} between each other, one of the levels can still form an \textit{EP} with some third level. As such,  the degenerate manifold of diabolical points is bounded by the manifolds of Exceptional Points.

In comparison, if $\zeta_1\zeta_2=\eta_1\eta_2$, then generally $\vec w$ is non-zero and is not collinear with $\vec u_-$. As a result, any non-zero $\Delta \vec p$ opens the gap, which is either real ($\zeta_1\zeta_2=1$) or imaginary ($\zeta_1\zeta_2=-1$).

\bigskip

\textit{Examples\/.} ---
We choose our examples from the class of $\BP\BT$-symmetric $\BP$-pseudo-Hermitian extensions of the transverse-field Ising model.
\begin{align}
\hat H &= \hat H_0 + \hat H_\mathrm{nh},\\
\hat H_0 & = \sum_{j=1}^N \Delta \hat\sigma^x_j - J\sum_{j=1}^{N-1} \hat\sigma^z_j\hat\sigma^z_{j+1},\\
\hat H_\mathrm{nh} & = i\sum_{j=1}^N\left(\gamma^z_j\hat \sigma^z_j + \gamma^x_j\hat \sigma^x_j\right).
\end{align}

Single qubits effectively described by non-Hermitian Hamiltonians
have already been experimentally realized in various solid state systems such as trapped ions, ultracold and Rydberg atoms \cite{ding2021experimental,lourencco2022non,li2019observation}, Bose–Einstein condensate \cite{cartarius2012model}, superconducting \cite{naghiloo2019,chen2021quantum,dogra2021quantum} or nitrogen-vacancies qubits \cite{wu2019observation}.
In principle, our exemplar class of models can be obtained in experiment by engineering the coupling between several such effective non-Hermitian qubits by coupling them, for example, to a lossless resonator~\cite{sfe-sw}.

We define the parity and the time reversal operators as the mirror flip of the chain and the complex conjugation respectively~\cite{Song-14,StarkovFistulEremin-22,StarkovFistulEremin-23}:
\begin{equation}
 \hat \BP \sigma_j^s\hat \BP^{-1} = \sigma_{N+1-j}^s,\quad \hat \BT i \hat \BT^{-1} = -i.
\end{equation}
The $\BP\BT$-symmetry of the model requires that
\begin{equation}
\gamma^z_j = -\gamma^z_{N+1-j},\qquad \gamma^x_j=-\gamma^x_{N+1-j}.
\end{equation}
In addition to $\BP\BT$-symmetric, the model is also pseudo-Hermitian with $\hat \zeta=\hat\BP$.
The Hermitian part $\hat H_0$ of the model is integrable and can be diagonalized via a combination of Jordan-Wigner and generalized Bogolyubov transformations~\cite{Lieb-61}. As such, one can compute the topological indices $\zeta_\BP$ corresponding to $\hat\BP$ analytically, which has already been done in Ref.~\cite{StarkovFistulEremin-23}.

We consider three specific arrangements of gain (loss) coefficients $\left\{\gamma_j^z,\gamma_j^x\right\}$.
\begin{itemize}
\item Purely longitudinal gain (loss): $\gamma_j^x\equiv0$. In this case, there is a second choice of pseudo-metric operator $\hat{\eta}=\hat \BU$, where
\begin{equation}
\hat \BU = \otimes_{j=1}^N \hat \sigma_j^x.
\end{equation}
The corresponding topological indices are $\eta = \zeta_\BU = -1^{n_{fm}}$, where $n_{fm}$ is the number of excited Bogolyubov fermionic modes. Note that the eigenspectrum is generally non-degenerate, so the corresponding topological indices are compatible.

\item Purely transversal gain (loss): $\gamma_j^z\equiv0$. In this case, $\hat \BU$ commutes with $\hat H$, and there is a second choice of the pseudo-metric operator $\hat {\eta}=\hat \BP\hat\BU$. The corresponding topological indices are $\eta=\zeta_\BP\zeta_\BU$.

\item Mixed gain (loss): $\gamma_j^z,\gamma_j^x\not\equiv0$. In this case, the symmetry is lowered, and there is no second choice of the pseudo-metric operator.
\end{itemize}

Dissipative dynamics described by a non-Hermitian Hamitlonian can be interpreted as a consequence of continuous measurements combined with the post-selection on measurement results~\cite{naghiloo2019}. In this context, tuning the combination of the gain (loss) parameters $(\gamma_i^x,\gamma_i^z)$ can be physically interpreted as changing the direction of the effective magnetic field used to probe the qubit $i$.

For concreteness, we assume $N=4$ and staggered gain (loss): $\gamma_j^z = \gamma\times(-1)^{j-1}$ in the purely longitudinal case, $\gamma_j^x=\gamma\times(-1)^{j-1}$ in the purely transversal case, and $\gamma_j^x=\gamma_j^z=0.5\gamma\times(-1)^{j-1}$. Using numerical diagonalization, we obtain the eigenspectra in each of the cases and plot them in Figs.~\ref{figz}, \ref{figx} and~\ref{figxz} respectively. There, we display the real parts of the normalized eigenvalues 
$\tilde\varepsilon = \varepsilon/\sqrt{J^2+\Delta^2}$ as the functions of the normalized Ising coupling $\tilde J = J/\sqrt{J^2+\Delta^2}$ for different fixed values of normalized gain (loss) strength $\tilde\gamma=\gamma/\sqrt{J^2+\Delta^2}$. In all of the figures, the imaginary parts of the eigenvalues are visualized as the shaded ribbons; the solid (dashed) linestyle denotes $\zeta=1$ ($\zeta=-1$). In Figs.~\ref{figz} and~\ref{figx}, the blue (orange) color of the lines denotes $\tilde \zeta=1$ ($\tilde \zeta=-1$).

In Figs.~\ref{figz} (longitudinal case) and~\ref{figx} (transversal case), we single out the crossings, where the levels have same indices with respect to one pseudo-metric, and opposite indices with respect to another one. 
According to the results of the previous section, these crossings form degeneracy lines in the two-dimensional parameter space $(\tilde J,\tilde{\gamma})$. As we change $\tilde\gamma$, such crossings do not open the gap and just shift in $\tilde J$, i.e. they are robust with the respect to the dissipation. At larger $\tilde\gamma$, a degeneracy line ends at a second-order \textit{EP}, formed by one of the crossing levels with some third level.

In Fig.~\ref{figxz}, we single out the same parity crossings.
Since there is no symmetry protection anymore, these crossings naturally open the gaps as soon as gain (loss) is turned on.\bigskip

\textit{Conclusions\/.} --- In this letter, we have focused on non-Hermitian systems with pseudo-Hermitian symmetry and studied the interplay of different parameter-independent choices of the pseudo-metric operator. We have shown it to be important both for the defective and non-defective degeneracies. Formation of a second-order~\textit{EP} between a pair of levels is possible only if it is allowed with respect to all pseudo-metric operators. Moreover, if the formation of a second-order~\textit{EP} between two levels is allowed for one choice of the pseudo-metric but forbidden for another one, the crossing of such levels has to be non-defective and is promoted to a whole manifold of non-defective degeneracies.

The results presented here are completely general and do not depend on the physical nature of the system realizing the non-Hermitian Hamiltonian as long as it has the symmetry properties we discussed. As such, the results can be also applied to the non-Hermitian Bloch Hamiltonians arising in topological band theory~\cite{Fu-18,Bergholtz-21} or to the effective non-Hermitian Hamitlonians arising in interacting many-body contexts~\cite{Kozii-17, Yoshida-18, Kazuhiro-19, Yoshida-20, Nagai-20, Michishita-20, Rausch-21, Lehmann-21, Crippa-21, Reitner-23}.\bigskip

\textit{Acknowledgements\/.} --- I would like to thank prof. Björn Trauzettel for the fruitful and encouraging discussions. 
This work was supported by the W{\"u}rzburg-Dresden Cluster of Excellence ct.qmat, EXC2147, project-id 390858490.


\bibliography{sising}

\end{document}